\documentclass[12pt,preprint]{aastex}

\newcommand{\bdv}[1]{\mbox{\boldmath$#1$}}

\def\au{{\rm AU}} 
\def\rel{{\rm rel}}
\def\e{{\rm E}}
\def\bpi{{\bdv\pi}}
\def\bmu{{\bdv\mu}}

\begin{document}
\title{{\it Kepler} Microlens Planets and Parallaxes}

\author{
Andrew Gould\altaffilmark{1},
Keith Horne\altaffilmark{2}
}
\altaffiltext{1}{Department of Astronomy, Ohio State University,
140 W.\ 18th Ave., Columbus, OH 43210, USA; 
gould@astronomy.ohio-state.edu}
\altaffiltext{2}{SUPA, University of St Andrews, School of 
Physics \& Astronomy, North Haugh, St Andrews, KY16 9SS, UK;
kdh1@st-andrews.ac.uk}

\begin{abstract} 
{\it Kepler}'s quest for other Earths need not end just yet: it
remains capable of characterizing cool Earth-mass planets by microlensing,
even given its degraded pointing control.  If {\it Kepler} were
pointed at the Galactic bulge, it could conduct a search for
microlensing planets that would be virtually non-overlapping with
ground-based surveys.  More important, by combining {\it Kepler}
observations with current ground-based surveys, one could measure the
``microlens parallax'' $\bpi_\e$ for a large fraction of the known
microlensing events.  Such parallax measurements would yield mass and
distance determinations for the great majority of microlensing planets,
enabling much more precise study of the planet distributions as functions
of planet and host mass, planet-host separation, and Galactic position 
(particularly bulge vs.\ disk).  In addition, rare systems (such as
planets orbiting brown dwarfs or black holes) that are presently lost
in the noise would be clearly identified.
In contrast to {\it Kepler}'s current primary hunting 
ground of close-in planets, its microlensing planets would be in the
cool outer parts of solar systems, generally beyond the snow line.
The same survey would yield a spectacular catalog of brown-dwarf
binaries, probe the stellar mass function in a unique way, and still
have plenty of time available for asteroseismology targets.
\end{abstract}


\section{{Introduction}
\label{sec:intro}}

The {\it Kepler} satellite has found more than 3000 planetary candidates,
the overwhelming majority of which are real planets 
\citep{batalha13}. To give one
example of the new parameter space probed, {\it Kepler} has discovered
231 ``Earth-radius'' planets (within 25\% of Earth's radius).  To date
{\it Kepler} has detected planets only by the transit method, and as
a result it is highly biased toward close-in planets.  For example,
the median period of the ``Earth-radius'' sample is 5.2 days, and the
maximum period is 69 days.

Here we propose to apply {\it Kepler} to characterizing much colder planets
in the outer parts of their solar systems, using the microlensing
technique \citep{gaudi12}.  We show that although {\it Kepler} is not optimally
designed for this task, it can be competitive with existing and
under-construction ground-based surveys in terms of finding
planets.

However, what {\it Kepler} would add that is fundamentally new
would be microlens parallaxes for a large fraction of microlensing events,
including almost all of those with planetary signals (whether
detected by {\it Kepler} or from the ground).  In the great majority
of cases, such parallax measurements would enable determination of
the host mass and distance, and thus also the planet mass, which would
greatly enhance the value of both groups of planets.

The requirements of a {\it Kepler} microlensing survey
are well-matched to the limitations on its performance due to loss of
pointing stability.  In order to be an effective transit-search tool,
{\it Kepler} had to monitor $\sim 10^5$ stars.  Given data-transmission
constraints, this implied relatively long (30 min) integrations on each
star, which in turn required high pointing stability.  However, to be
an effective microlensing-planet tool, it need only observe $\sim 10^3$
stars.  Hence the same data-transmission constraints are compatible with
much shorter exposures.

The photometric requirements of microlensing planet searches are very different
from {\it Kepler}'s transit survey.  Planetary deviations are typically
tens of percent, compared to $\la 1\%$ for transits.  However, the source
stars are much fainter, typically $18\la I\la 16$, compared to $V\la 16$
for the transit survey.  Microlensing events typically last a few weeks to
months. They are usually quickly identified from the ground, but {\it Kepler}
would have to be notified of these identification to conduct its search.
Planetary deviations due to Jupiter-mass planets typically last one day, while
those due to Earth-mass planets typically last about one hour.  Hence somewhat
shorter cadences are needed than {\it Kepler}'s traditional 30 min in order
to get full sensitivity to the lowest-mass planets.

The photometric requirements for microlensing parallax measurements are
substantially less restrictive than for finding planets because the
parallax signal extends over the entire event, not just a few hours or days.
This is important: it means that even if the photometric challenges 
prove too difficult
to find a large number of planets on its own, {\it Kepler}'s main
contribution of precise characterization of ground-based planets
can remain intact.

\section{{Observation Strategy}
\label{sec:strat}}

At present, roughly 2000 microlensing events are discovered per year
by the Optical Gravitational Lens Experiment 
(OGLE\footnote{http://ogle.astrouw.edu.pl/ogle4/ews/ews.html/}) 
and Microlensing
Observations for Astrophysics 
(MOA\footnote{https://it019909.massey.ac.nz/moa/}) 
collaborations.  The overwhelming 
majority of these are found in a region that could fit in a single
pointing of the $105\,\rm deg^2$ {\it Kepler} camera.  Thus, the first
element of the strategy would be simply to point {\it Kepler} at this
field, when permitted by its $55^\circ$ Sun exclusion angle.  Whenever a
new microlensing event was found (from Earth), it would be added to
the list of {\it Kepler} targets.  Most events are detected at least
several days before they do anything interesting, so such ``uploads''
of new targets could be grouped in batches, if necessary.  Microlensing 
events could also be removed from the list when they returned to baseline.

We note that {\it Kepler} is in a $P=372.5\,$d orbit and so drifting
behind Earth at $7.2^\circ\,\rm yr^{-1}$ and hence is now roughly
1 month (0.5 AU) behind Earth.  This is an excellent position to
create a large baseline for ``parallactic viewing'' while still having
a strongly overlapping ``bulge season'' with Earth.

\section{{Unique Impact: Microlensing Parallaxes}
\label{sec:parallaxes}}

The observational strategy outlined above would accomplish two aims:
measure the ``microlens parallax'' of a large fraction of events
and detect planets in a subset.  We will argue below that the planet-finding
capability is comparable but not qualitatively superior to ground-based
capabilities.  Hence, we focus first on what is unique about a {\it Kepler}
microlensing survey: parallaxes.

\subsection{{What is microlensing parallax, $\bpi_\e$?}
\label{sec:what}}

The magnitude of the microlens parallax, $\bpi_\e$ is simply the lens-source
relative parallax, $\pi_\rel$, scaled to the angular \citet{einstein36} 
radius $\theta_\e$
\begin{equation}
\pi_\e = {\pi_\rel\over\theta_\e};
\quad
\theta_\e^2=\kappa M \pi_\rel;
\quad
\kappa\equiv {4 G\over c^2 {\rm AU}} = 8.1 {{\rm mas}\over M_\odot}
\label{eqn:piedef}
\end{equation}
\citep{gould92,gould04},
while its direction is that of the lens-source relative proper motion,
$\bpi_\e/\pi_\e = \bmu/\mu$.

\subsection{{Parallax: Rosetta Stone for microlensing planets}
\label{sec:why}}

The significance of a parallax measurement is that if $\theta_\e$
is also measured, then one can immediately derive $M$ and $\pi_\rel$,
\begin{equation}
M = {\theta_\e\over \kappa \pi_\e},
\quad
\pi_\rel = {\au\over D_L} - {\au\over D_S} = \theta_\e\pi_\e,
\label{eqn:massdis}
\end{equation}
Since the source distance $D_S$
is usually known quite well, measuring $\pi_\rel$
immediately gives the lens distance $D_L$.

While in general it is quite difficult to measure $\theta_\e$, such
measurements are
almost always possible in planetary lensing events.  This is because
the source must pass over or near a ``caustic'' caused by the planet
if the planet is to be detected.  The lightcurve deviation is therefore
a function of $\rho\equiv \theta_*/\theta_\e$, where $\theta_*$ is the 
angular source size, which means that $\rho$ can almost always be measured
from the lightcurve of planetary events.  Since $\theta_*$
can be routinely measured from the source color and magnitude
\citep{yoo04}, $\theta_\e=\theta_*/\rho$ can also be measured.

Hence, microlens parallax is a Rosetta Stone for planetary microlensing events,
turning what was initially thought to be a purely statistical technique
\citep{gouldloeb} into individual planet-mass and distance measurements.

\subsection{{How is microlens parallax measured?}
\label{sec:how}}

To date, the overwhelming majority of microlens parallax measurements
have relied on observing lightcurve deviations induced by the accelerated
motion of Earth during the event \citep{gould92,alcock95,poindexter05}.
However, because most microlensing events are short compared to the
time required for Earth to move a radian (yr/$2\pi\sim 58\,$d), such
``orbital'' parallax measurements are quite rare.  Another approach is
to simultaneously observe the event from two locations on Earth
\citep{hardy95,holz96}, but since the projected Einstein radius
$\tilde r_\e\equiv \au/\pi_\e$ is typically several AU, this is only
practical for extreme magnification events $A\ga 1000$ \citep{gould97},
and in fact there are only two such cases \citep{ob07224,ob08279}.

Therefore, the only method that can {\it routinely} return microlens
parallaxes is to combine observations from a satellite at $\cal{O}(\au)$
from Earth and so enable simultaneous observations from two locations
separated by a distance that is comparable to $\tilde r_\e$
\citep{refsdal66,dong07}.  

{\it Kepler} therefore possesses two
tremendous advantages for a microlens parallax survey: it is already
in solar orbit and it can observe essentially all ongoing microlensing
events simultaneously.

\subsection{{Parallax degeneracies}
\label{sec:degen}}

\begin{figure}
\includegraphics[scale=0.72]{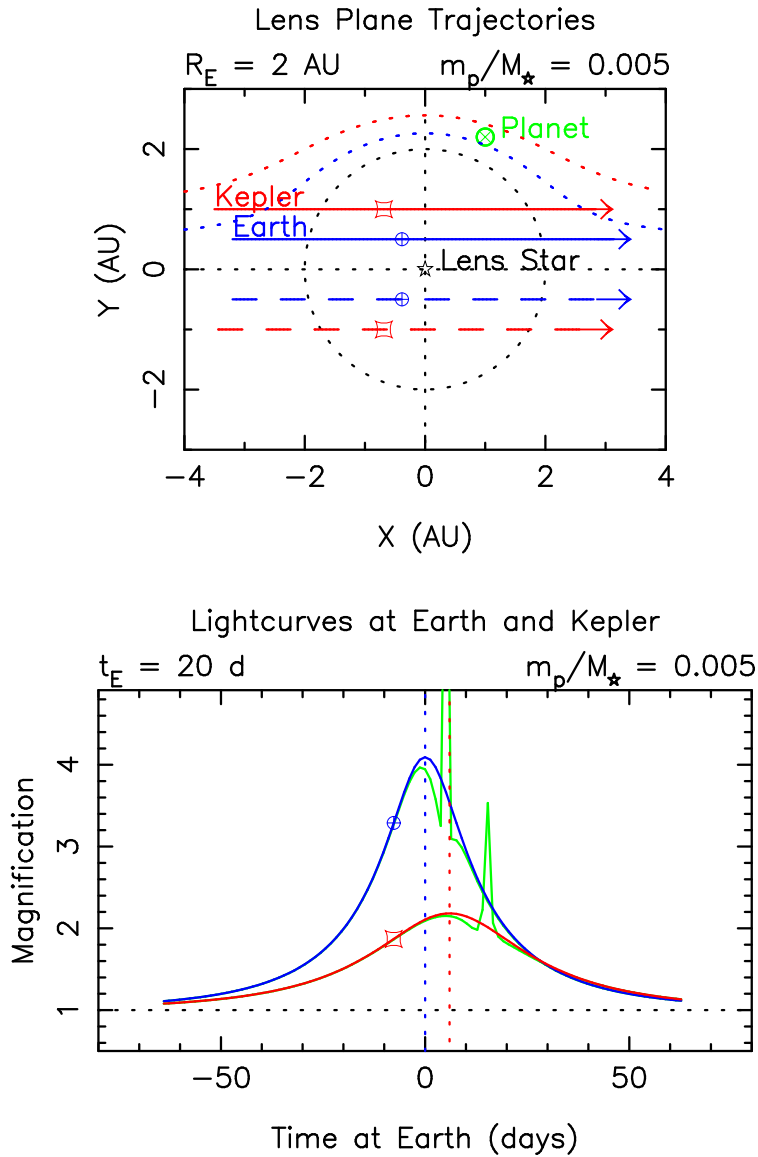}
\caption{\label{fig:geom}
Illustration of four-fold degeneracy derived from comparison of {\it Kepler}
and ground based lightcurves.  Upper panel shows two possible trajectories
of the source relative to the lens for each of {\it Kepler} (red) and Earth 
(blue) observatories.  Each set would give rise to the same point-lens 
lightcurve in the lower panel (same colors), leading to an ambiguity
in the Earth-{\it Kepler} separation (distance between red circle and blue
square) relative to the Einstein ring.  In this particular case, the planet
causes deviations to both lightcurves (green), thus proving that the
trajectories are on the same side of the Einstein ring.  More generally,
the planet would appear in only one curve, leaving the ambiguity open.
In this case, it would be resolved by more subtle differences in the
Einstein timescale.  See \citet{gould94,gould95}.
}
\end{figure}

However, it also faces challenges.  Some of these are
related to its relatively large point spread function (PSF), which
we will discuss below.   But one challenge is rooted in the nature of 
space-based parallax measurements: degeneracy.  As already noted
by \citet{refsdal66} and discussed more thoroughly by \citet{gould94},
space-based parallax measurements are subject to a four-fold discrete
degeneracy.  Basically, Earth and satellite see the same
event, but displaced in the Einstein ring, and so having different peak times
$t_0$ and different impact parameters $u_0$.  The microlens parallax,
is then given essentially by
\begin{equation}
\bpi_\e = {\au\over D_{\perp,\rm sat}}(\Delta\tau,\Delta\beta);
\qquad
\Delta\tau\equiv {t_{0,\rm sat}-t_{0,\oplus}\over t_\e};
\qquad \Delta\beta \equiv u_{0,\rm sat}-u_{0,\oplus},
\label{eqn:satpar}
\end{equation}
where $D_{\perp,\rm sat}$ is the Earth-satellite separation
projection onto the plane of 
the sky and $t_\e$ is the Einstein timescale.  The problem is that while
$t_0$ is uniquely determined from the lightcurve, $u_0$ is a signed quantity
whose magnitude is measured but not its sign.  Thus, $\bpi_\e$
can take on four values depending on the signs of $u_0$ as seen from
Earth and the satellite.  
However, since the mass depends only on the
magnitude of $\bpi_\e$, only a two-fold degeneracy is really of major
interest.  That is, do $u_{0,\rm sat}$ and $u_{0,\oplus}$ have the same or
opposite signs?  Or, equivalently: is the source seen projected on the
same or opposite side of the lens as seen from the two observatories?
See Figure~\ref{fig:geom}, and also Figures 1 and 2 from \citet{gould94}.

\citet{gould95} showed that this degeneracy could be broken because
the timescales of the events as seen from Earth and the satellite are
slightly different, and this difference is a function of $\Delta\beta$.
\citet{gaudi97} then investigated how well this degeneracy could be
broken for events seen toward the Galactic Bulge.  Their assumptions
were far more conservative than those likely to apply to {\it Kepler}
observations.  First, they considered a narrow-angle pointed mission
(rather than a wide-angle survey), in which the
observations of each target would be limited to a relatively few epochs,
whereas {\it Kepler} observations would be continuous.  Second, at
the time it was believed that the source flux in the space-filter
could not be accurately determined from the ground-based lightcurve,
whereas subsequently \citet{mb11293} have shown that this indeed is possible
to at least 1\% precision.  See also \citet{gould13} and \citet{yee13}.

Undoubtedly, there will be microlensing events discovered that are so faint
that the parallax degeneracy will not be broken.  However, few planets
are likely to be found in such faint events.  
Moreover, depending on the geometry
of the event, it is sometimes not necessary to actually break the degeneracy
to derive good mass estimates (e.g., if $|\Delta\tau|\gg |\Delta\beta|$).

\section{{Planetary Science with {\it Kepler} Microlens Parallaxes}
\label{sec:science}}

At present, most microlensing planet detections return $\theta_\e$ and hence
the product $M\pi_\rel = \theta_\e^2/\kappa$, but not the mass and distance
separately.  Hence, these quantities are estimated only statistically
for most events.  The estimates make use of Galactic models together
with various pieces of information, such as the geocentric lens-source
relative proper motion $\mu=\theta_\e/t_\e$ and upper limits on the lens
flux from blended light.  But generally these estimates are accurate to
only a factor of two, and of course can be radically incorrect in cases
of unusual or unexpected systems.  In particular, there is only one
planet out of about 30 detected to date that is known to be in the Galactic
bulge with good confidence, even though the majority of lenses are in
the bulge.  Hence, it is very difficult to disentangle the distributions
of planets as functions of controlling properties, such as planet mass,
host mass, distance from host, and Galactic position.

In one fell swoop, {\it Kepler} could resolve all of these uncertainties,
and it could do so for the several dozen planets per year that will
be discovered with current and in-construction experiments.  For example,
standard core-accretion theory predicts a dip in the planet mass function
between Neptunes and Jupiters.  By sharpening the mass resolution
of microlens planets, {\it Kepler} could directly test this prediction
for ice and gas giants found beyond the snow line, which is presumably
their birth place.

Not only would this increased precision be of direct use in better
understanding the planets that microlensing is discovering, it would
also put them ``on the same playing field'' as the planets, 
mostly much closer to their hosts, discovered by other techniques.

\section{{{\it Kepler} Cold Planets}
\label{sec:cold}}

In addition to measuring microlens parallaxes, {\it Kepler} observations
will probe a virtually independent set of microlensing planets from those
detected from the ground.  This is because it is displaced from Earth
by $D_{\perp,\rm sat}/\tilde r_\e$ in the Einstein ring, which will typically be
of order 10\%.  Since planetary perturbations are usually much smaller
than this, planets detected from Earth will generally not be detected
by {\it Kepler} and vice-versa.  

Here, we estimate the general competitiveness
of {\it Kepler} relative to the Korea Microlensing Telescope Network 
(KMTNet, \citealt{kmtnet})
which is the largest ground-based microlensing experiment currently under
construction.  KMTNet will have three 1.6m telescopes, each with a 
$4\,\rm deg^2$ field of view, located in Chile, South Africa, and Australia.
Like {\it Kepler}, therefore, it is in principle capable of near-continuous
coverage for the fraction of the year when the Sun is well away from the
bulge.  KMTNet will cycle through four fields, observing each for 2 min
out of 10.  Considering weather at these sites, it will have a combined
duty cycle of perhaps 2/3.  In addition, {\it Kepler} has a ``white-light''
response compared to $I$-band filters that are needed from the ground.
Taking account of the increase in both signal and noise implies a factor
1.6 advantage.  Altogether, these factors give {\it Kepler} an advantage
of a factor 12.

However, {\it Kepler} has disadvantages as well, and these are overall
stronger.  First, its aperture is smaller by a factor $1.6^2$.  Most
important, its PSF is much larger.  At best, the FWHM $\sim
3.1^{\prime\prime}$, whereas average KMTNet seeing is likely to be
$\sim 1.2^{\prime\prime}$.  Since almost all photometry is likely to
be below sky in either case, these two disadvantages combine to a
factor $(1.6\times 3.1/1.2)^2 \sim 17$.  Finally, the problems posed
by field drift are difficult to estimate without detailed simulations.
The exposures can be made short enough that this drift does not affect
individual images, but the undersampled PSF, in conditions of crowded
bulge fields is likely to increase the photometric noise beyond the
above naive calculation.  Thus, {\it Kepler} will find fewer planets
within the $16\,\rm deg^2$ probed by KMTNet, which contain the richest
planet hunting ground.  By the same token, of course, a {\it Kepler}
microlens planet search would fall far short of one by {\it WFIRST}
\citep{green12,spergel13}.
However, {\it Kepler} will also find planets in outlying
regions, which are being surveyed by OGLE and MOA at lower cadence.
And, as emphasized above, virtually all the planets that it does find
will be undetected from the ground.

Nevertheless, this calculation shows that the planets found by {\it Kepler}
are not reason enough, by themselves, for it to do a microlensing survey.
Moreover, since it will be looking only at events found by others, it will
be useless for finding free-floating planets (FFP), 
which are a unique capability
of microlensing\footnote{It will, however, be useful for vetting the main
contaminant of the FFP signal, stellar microlensing events whose
timescales are exceptionally short due to small $\pi_\rel$ despite high mass
$M$.
If FFP events can be alerted to {\it Kepler} within $\sim 1\,$d,
they will appear similarly for {\it Kepler} because their parallaxes will
be small $\pi_\e=(\pi_\rel/\kappa M)^{1/2}$.  However, if they are due to
planets, then {\it Kepler} will see no event at all because the large
parallax puts the source well outside the Einstein ring from {\it Kepler}'s 
perspective}
\citep{sumi11}.
Rather, its principal value is to obtain microlens parallaxes, which
would enormously enhance the value of planets detected in ground-based
surveys.

\section{{Other Microlensing Applications}
\label{sec:otherulens}}

Microlensing surveys are also a powerful probe of binaries, in particular
low-mass binaries that are difficult or impossible to detect by other
methods.  For example, \cite{choi13} discovered two
brown dwarf binaries, which obeyed the binding-energy floor found previously
using standard brown-dwarf search techniques, but at much lower mass and
tighter separation.  That these binaries yielded mass measurements
(and so could even be recognized as brown dwarfs, not stars) was only due
to the fact that they were unusually nearby (few kpc) and so had large,
easily measurable parallaxes.  Like planetary events, binary events
routinely yield $\theta_\e$, so that {\it Kepler} microlens parallaxes would
give masses and distances for all binaries, and so sift out these brown-dwarf
binaries, which are otherwise generally unrecognizable.  Moreover, 
binaries, in contrast to planets, would often be detected by both {\it Kepler} 
and ground observatories, which would provide detailed information on their
orbits.  Finally, while the point-lens events would not generally yield
$\theta_\e$ (and so masses), their mass function could be studied statistically
from a {\it Kepler} microlens parallax survey \citep{han95}.

\section{{Non-microlensing Applications}
\label{sec:nonulens}}

The number of microlensing targets to be observed is not large,
at most 2000 in a season, and not all must be observed all season.  The
exposure times must be fairly short because of {\it Kepler}'s degraded
pointing stability, but it is unlikely that they need to be 100 times
shorter than {\it Kepler}'s traditional 30 min exposures.  From a microlensing
standpoint, there are no drivers for exposures shorter than about 5 minutes.
Thus, it is likely that microlensing targets will absorb only a small
fraction of the available data-transmission capability.  Other objects,
such as bright asteroseismology targets could therefore
be observed.  In particular,
since short exposures are needed due to stability problems, one could
target bright dwarfs, which have higher-frequency oscillations than
the giant-star targets on which {\it Kepler} has concentrated to date.

\acknowledgments

We thank Scott Gaudi and Jennifer Yee for stimulating discussions.
Work by AG was supported by NSF grant AST 1103471 
and NASA grant NNX12AB99G.  KH is supported by a Royal 
Society Leverhulme Trust Research Fellowship.

\end{document}